# Experimental observation of chimera states in spiking neural networks based on degenerate optical parametric oscillators

**Authors:** Tumi Makinwa[1,2†], Kensuke Inaba[1*†], Takahiro Inagaki[1*†], Yasuhiro Yamada[1], Timothée Leleu[3], Toshimori Honjo[1], Takuya Ikuta[1], Koji Enbutsu[4], Takeshi Umeki[4], Ryoichi Kasahara[4], Kazuyuki Aihara[3] & Hiroki Takesue[1]

[†]These authors contributed equally to this work.

**Affiliations:** [1] NTT Basic Research Laboratories, NTT Corporation, 3-1 Morinosato Wakamiya, Atsugi, Kanagawa, 243-0198, Japan

[2] Department of Applied Sciences, Delft University of technology, Mekelweg 5, 2628 CD Delft, Netherlands

[3] International Research Center for Neurointelligence, The University of Tokyo Institute for Advanced Study, The University of Tokyo, 7-3-1 Hongo, Bunkyo-ku, Tokyo 113-0033, Japan

[4] NTT Device Technology Laboratories, NTT Corporation, 3-1 Morinosato Wakamiya, Atsugi, Kanagawa, 243-0198, Japan

*E-mail to: Kensuke Inaba (kensuke.inaba.yg@hco.ntt.co.jp) and Takahiro Inagaki (takahiro.inagaki.vn@hco.ntt.co.jp)

**Abstract**

We experimentally demonstrate that networks of identical photonic spiking neurons based on coupled degenerate parametric oscillators can show various chimera states, in which, depending on their local synchronization and desynchronization, different kinds of spiking dynamics can develop in a self-organized manner. Even when only a static interaction is implemented, through synchronized inputs from connected neurons, the spiking mode of photonic neurons can be spontaneously and adaptively changed between the Class-I and Class-II modes classified by A. L. Hodgkin. This spontaneous spiking-mode shift induces a significant change in the spiking frequency despite the all neurons having the same natural spiking frequency, which encourages the generation of chimera states. Controllability and self-organized flexibility of the spiking modes in the present system allow us to create an experimental platform to explore the nature of chimera states in neuromorphic spiking neural networks.

# 1. Introduction
## 1.1 Background

Chimera states in networks of coupled oscillators have been the subject of considerable interest in complex nonlinear systems [1,2]. Kuramoto et al. pointed out the counterintuitive fact that identical oscillators with homogeneous interactions can achieve an inhomogeneous state in which the phases of some of them are in synchronization, while those of the others are out of synchronization [1]. This 'mixed' state of synchronized and desynchronized oscillators has been named the chimera state [2]. In fact, chimera-like behavior has been observed in the real world, such as in a national power grid [3] and in condensed matter [4,5]. Neuromorphic and biological systems also show chimera-like phenomena, including social systems [6,7], neuronal bump states [8], contractions of heart cells [9], unihemispheric sleep [10], and diseases related to the brain [11-13]. In response to these findings, the trend of theoretical studies on neural networks have been changing to spiking neural network (SNN) models [14-17] composed of the Fitzhugh-Nagumo [18-20], Hindmarsh-Rose [21-23], Morris-Lecar [24], Wilson-Cowan [25], and other models [26,27].

It was initially thought that the occurrence of chimera states requires a very strict set of conditions to be imposed on the networked oscillators, such as in the Kuramoto model [17]. For example, it was believed that non-local interactions were required to induce chimera states. However, recent research on SNNs, e.g., with the Hindmarsh-Rose model [21], has shown that they can occur under a broader range of conditions than believed prior [17], e.g., local (nearest-neighbor) [28] or global (all-to-all) interactions [29]. Furthermore, it has been proposed that adaptive interactions, namely, dynamically changing ones such as biological systems with Hebbian rules, stabilize chimera states in a self-organized manner [19]. In addition, the discovery of amplitude chimeras has led to a broadening of the definition of chimera states [30], and a scheme to phenomenologically categorize chimeras has been established that can be widely applied to simple oscillatory and SNN systems as well as to experimental results [17]. However, while theoretical studies on chimera states in various systems, in particular, SNN systems [14-27], have become widespread, so far, there has been little experimental research on chimera states in SNN systems [31], even though various oscillatory networks [32-35] and chaotic optical systems [36-39] have been experimentally studied. Experiments on SNNs would allow us to simulate real-world chimera-like phenomena [6-13] and may lead to progress on, for example, understanding brain diseases.

In this paper, we report on various chimera states that were experimentally observed in a photonic SNN through control of the experimental parameters and propose an experimental platform to research chimera states in the SNN. Our neuromorphic neurons each consist of a pair of degenerate optical parametric oscillators (DOPOs). Pairs of

DOPOs can exhibit two spiking modes from Hodgkin's classification [40], which allows our optical system to mimic neuronal dynamics in realistic biological systems. Furthermore, synchronization of the networked neurons causes a spontaneous spiking-mode shift accompanied with a large frequency change [41]. This unique property stabilizes the chimera states and causes their remarkable moving behavior.

## 2. Theory
### 2.1 DOPO spiking neuron

A photonic SNN can be represented as a network of paired DOPOs. Each single DOPO is a nonlinear oscillator generated by using a phase sensitive amplifier (PSA) based on degenerate parametric amplification in a cavity. A DOPO can oscillate only in two phase states of 0 or π above the threshold pump power of the oscillator [42,43]. Optical coherence between these bistable states is essential to represent the membrane potential of the neuron. By introducing energy transfer between paired DOPOs, named $v$- and $w$-DOPOs, through antisymmetric coupling, as shown in Fig. 1(a), the spiking dynamics can be implemented with optical amplitudes $v_i$ and $w_i$ of the $i$ th pair of DOPOs that can be described as follows:

$$\frac{dv_i}{dt} = (-1 + p_i)v_i - v_i^3 + J_{vw}w_i + \alpha \sum_j J_{ij}v_j - \beta \sum_j J_{ij}w_j \tag{1}$$

$$\frac{dw_i}{dt} = (-1 + p_i)w_i - w_i^3 + J_{wv}v_i + \alpha \sum_j J_{ij}w_j + \beta \sum_j J_{ij}v_j \tag{2}$$

where $p_i$ is the optical pump amplitude normalized by the oscillation threshold of each DOPO, and $J_{vw}$ and $J_{wv}$ with $J_{vw}J_{wv} < 0$ are antisymmetric intra-neuron coupling strengths. Here, we introduce parallel and cross-type inter-neuron interactions with weights $\alpha$ and $\beta$, as shown in Fig. 1(b) and (c). Without these interactions ($\alpha = \beta = 0$), by changing the operating parameters consisting of the pump amplitudes $P_i = -1 + p_i$ and basic frequency $\omega_0 = \sqrt{-J_{vw}J_{wv}}$, a single DOPO neuron exhibits the spiking modes of Class-I and Class-II neurons, as originally classified by A. L. Hodgkin [40]. One of the differences between these classes is that Class-I (-II) neurons show a (dis-) continuous change in spiking frequency over a wide (narrow) range as a parameter $P_i/\omega_0$ changes (see the Supplementary Information).

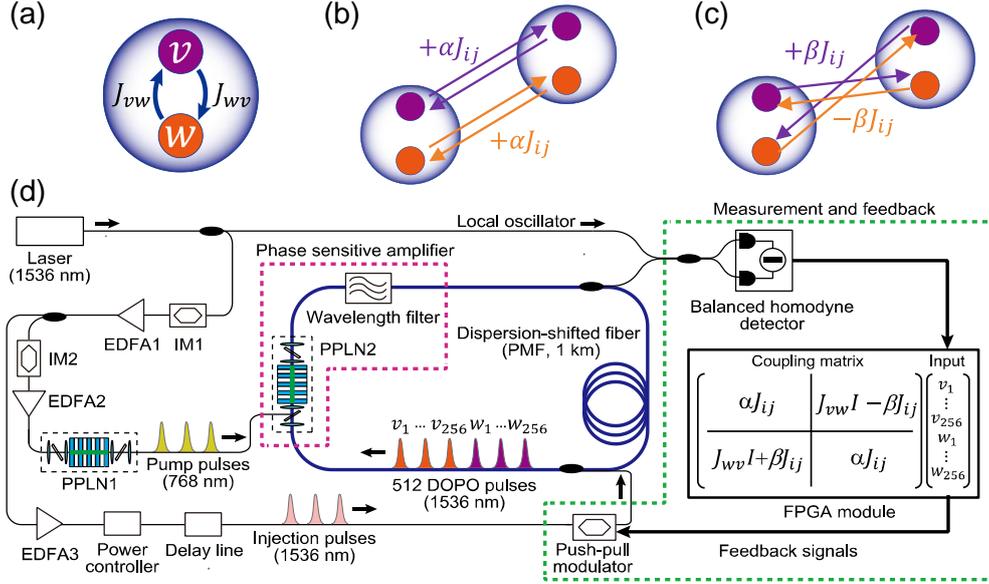

**Figure 1: Experimental setup of photonic spiking neural network** (a) Single photonic neuron composed of coupled degenerate optical parametric oscillators (DOPOs). (b) Parallel-type coupling between $i$ and $j$th DOPO neurons. (c) Cross-type coupling. (d) Schematic diagram of a photonic spiking neural network with 256 DOPO neurons. PPLN, periodically poled lithium-niobate; PMF, polarization-maintaining fiber; IM, intensity modulator; EDFA, erbium-doped-fiber amplifier; FPGA, field-programmable gate array.

## 2.2 Spontaneous mode change

Our previous study with parallel-type couplings ($\alpha = 1$ and $\beta = 0$) showed that the spiking modes of synchronized neurons spontaneously shift from Class-II to Class-I [41]. In the present study, we used cross-type coupling with $\alpha = 0$ and $\beta = 1$, which induces a spiking-mode shift from Class-I to Class-II after synchronization, in contrast to the parallel-type coupling case. This behavior can be understood from the fact that the order parameter of the synchronization effectively renormalizes the pump $P_i$ and basic frequency $\omega_0$ for parallel and cross-type interactions, respectively (see Method). Note that the renormalized parameter $P/\omega_0$ effectively determines the class of the synchronized neurons, where a small (large) $P/\omega_0$ induces Class-II (-I) type neurons. This function of the self-organized spiking-mode shift can assist various synchronous phenomena, because the spiking frequencies can be adjusted.

## 2.3 Experimental setup

Our photonic spiking neural network consists of a 1-km fiber ring cavity, a PSA module, and a measurement and feedback (MFB) system, as shown in Fig. 1(d) [41,44]. A DOPO pulse is generated by degenerate parametric amplification, which is provided by the PSA module based on $\chi_2$ nonlinearity of periodically poled lithium-niobate (PPLN) waveguides in the cavity [45]. An optical pulse at a wavelength of 1536 nm is converted into a pump pulse at a wavelength of 768 nm through second harmonic generation in one of the PPLN waveguides (PPLN1). By sending the converted pump pulse into the other PPLN waveguide (PPLN2) in the cavity and setting an optical band-pass filter that has a passband width of 13 GHz and center wavelength of 1536 nm, the DOPO pulse transmitting through the PSA module undergoes degenerate parametric amplification. When the pump amplitude increases above an oscillation threshold, the optical output of the DOPO bifurcates to a 0 or π phase state. To generate a large number of DOPO pulses in the 1-km fiber ring cavity, sequential pump pulses with a 60-ps width and 1-GHz repetition frequency are input to the PSA module. The round-trip time of the cavity is about 5 μs, and thus, 5,000 DOPO pulses can be generated with time-domain multiplexing. Arbitrary network structures can be implemented with the MFB system, in which optical couplings between DOPO pulses are achieved by injecting optical feedback pulses into the cavity. In this system, the in-phase component of each DOPO output is measured by a balanced homodyne detector during each round trip. These measured signals are multiplied with a coupling matrix by a field-programmable gate array (FPGA) module. The calculated results of the FPGA module are then converted into optical phases and amplitudes of the optical feedback pulses by an optical modulator. By repeating this procedure in each round trip, DOPO pulses in different time slots can behave as a network. The 512 DOPO pulses can be networked with an arbitrary coupling matrix composed of signed 8-bit integers. Since a single DOPO neuron consists of a pair of DOPOs, this experimental setup can implement photonic SNNs with 256 DOPO neurons, where 56 neurons are used as references to check the experimental conditions.

## 3. Results
### 3.1 Stationary chimera state with dense network

We investigated the chimera states on a ring lattice structure composed of 100 neurons. The neurons within the ring were made to be identical by setting $P_i = P$ for all $i$. First, we used a constant long-range interaction $J_{ij} = K\Gamma(r)$ with a distance $r = |i - j|$ and $\Gamma(r) = 1$ for $0 < r \leq d_{max}$ and 0 otherwise, as illustrated in Fig. 2(a). Here, we set the interaction length $d_{max} = 35$, pump amplitude $P/\omega_0 = 2.4$, and coupling strength $K/\omega_0 = 0.067$. Figure 2(b) shows the dynamics of the phases of the neurons, where $\theta_i = \tan^{-1}(w_i/v_i)$, and Fig. 2(d) shows a snapshot at 20.48 ms in Fig. 2(b). These results indicate the coexistence of two kinds of region, synchronized and desynchronized. We experimentally found a double-headed chimera state, which is analogous to the results obtained from numerical calculations based on the Fitzhugh-Nagumo model [18]. Note that changing the parameters, $d_{max}$, $K$, and $P$, complexly affect the properties of the chimera, such as, the number and size of the heads, which will be discussed elsewhere.

Here we focus on the characteristic meandering movement such as that of "snake", which was not observed in the spatial-stable chimeras in Ref. [18]. This meandering movement shows similar features with alternating chimera [20] (See the Supplementary Information).

To clarify the properties of this chimera, we calculated the local curvature measure $D$ introduced in Ref. [17], which is the second derivative of the phase of a neuron and its nearest neighbors: $D = f(x - \Delta x) - 2f(x) + f(x + \Delta x)$ with $f(x) = \theta_i$ and $\Delta x = 1$. The local curvature presented in Fig. 2c clearly distinguishes the regions: one is synchronized with $D \sim 0$ and the other is desynchronized with $D \neq 0$. In addition, we calculated the probability density $g$ as a function of $|D|$, where the density of the synchronized components, defined as $g_0(t) = \int_0^\delta g(t, |D|) d|D|$, represents the relative size of the synchronized regime at time $t$ [17]. We set the threshold to $\delta = 0.1 D_{max}$, where $D_{max}$ is the maximum value of $|D|$. The measure $g_0(t)$ can identify the chimera state when $0 < g_0(t) < 1$, where $g_0(t) = 0$ and $1$ represent asynchronous and synchronous states, respectively. In addition, the time dependence of $g_0(t)$ characterizes the spatial fluctuations of the chimera states [17]. We also analyze density function $h$ of $|\rho_{ij}|$ by calculating quantity $\rho_{ij} = \frac{\langle (Z_i - \mu_i)^*(Z_j - \mu_j) \rangle}{\sigma_i \sigma_j}$, where $\langle \cdot \rangle$ is the time average, $Z_i = \exp i\theta_i$, $\mu_i$ and $\sigma_i$ are means and standard deviations of $Z_i$, respectively [17]. Here, the density of the temporary stable component $h_0$, defined as $\sqrt{\int_\gamma^1 h(|\rho|) d|\rho|}$ with $\gamma = 0.9$, characterizes the temporal fluctuations of the chimera states. By using these indices $g_0(t)$ and $h_0$, the chimera state can be well categorized as follows [17]. Figure 2e shows $g_0(t)$ to be almost constant except for some experimental fluctuations, and shows $h_0$ to be nearly equal to 0, suggesting that the observed chimera state is categorized as a moving-stationary chimera state. Here, "stationary" means that the ratio of the spatial regions of chimera states stays almost constant, and "moving" comes from meandering movement of the chimera.

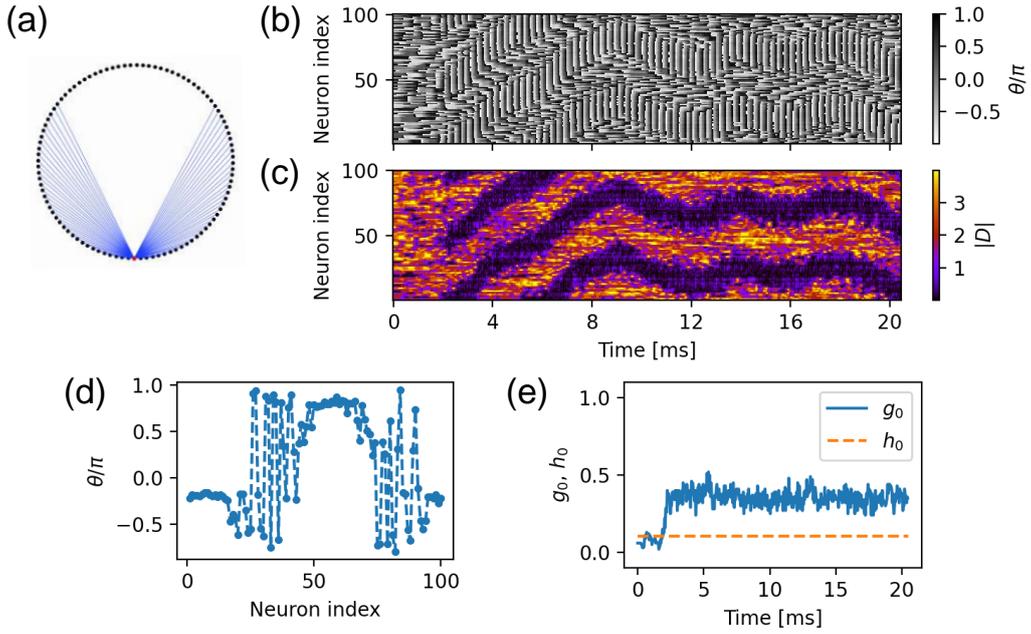

Figure 2: Experimental results for flat network ($d_{max}$ = 35) (a) Network structure of each neuron. (b) Rotation phases of neurons. (c) Local curvature $D$. (d) Phase distribution at t=20.48 ms. (e) $g_0(t)$ and $h_0$ indices. Pump power and coupling strength are set to be $P/\omega_0 = 2.4$ and $K/\omega_0 = 0.067$, respectively.

### 3.2 Spiking mode change by synchronization

Here, we discuss the cause of the chimera state meandering. Note that this meandering movement does not originate from noise in the experiments, because the numerical simulations without noise reproduce this movement (see the Supplementary Information). Rather, this behavior originates from the spontaneous change of the spiking modes at the boundary between the desynchronized and synchronized regions. Figures 3(a) and (b) show trajectories of specific neurons in the $v$-$w$ plane and the dynamics of the neuron phase $\theta_i$, where the left, middle, and right panels respectively show these quantities for neurons #40-45 in the desynchronized region, #65-70 in the synchronized region, and #30-35 at the boundary between the two types of regions. Interestingly, the identical neurons clearly show different spiking modes, namely, the stepwise $\theta$ of Class-I (desynchronized region) and the linear $\theta$ of Class-II (synchronized region). We found that changes in class occasionally occurs at the boundary. The synchronization causes a spontaneous spiking-mode shift from Class-I to Class-II as mentioned above (see Method). These two classes have clearly different spiking frequencies, as shown in Fig. 3(c). The main characteristic of the Class-I spiking mode is the continuous change in spiking frequency from zero. This results in meandering movements, because a Class-I neuron at the boundary between synchronized and desynchronized regions can adjust its spike timing (see Fig. 3(b) right panel) and then chaotically go back and forth between regions depending on the randomness coming from the initial state. This meandering movement occurs even without noise (see the Supplementary Information).

It should be noted that the present system has inversion symmetry between $v$ and $w$, meaning that the time scales of two-dimensional variables are intrinsically equivalent to each other. This fact suggests that the slow-fast multiple-time scales, which are usually included in the SNN systems, such as, fast-electronic and slow-chemical synaptic dynamics, is not a prerequisite for chimeras to appear in the SNN [14]. However, in the present system, the change in the spiking mode spontaneously causes a large time-scale difference between synchronized Class-II neurons with high spiking frequencies and desynchronized Class-I neurons with very low frequencies, which may stabilize the chimera states. A similar self-organized mechanism for stabilizing chimeras has been discussed in Ref [19], which considers adaptive interactions, namely, dynamically changing interaction strengths, while the present system only has static interactions. Note that the present system intrinsically contains adaptive controllability of spiking frequencies via synchronization [41]. In addition to causing the frequency difference, the spiking-mode shift changes the amplitudes in each region, which may also stabilize the chimera states, as in the case of the amplitude chimera found in Ref. [30] (see the Supplementary Information).

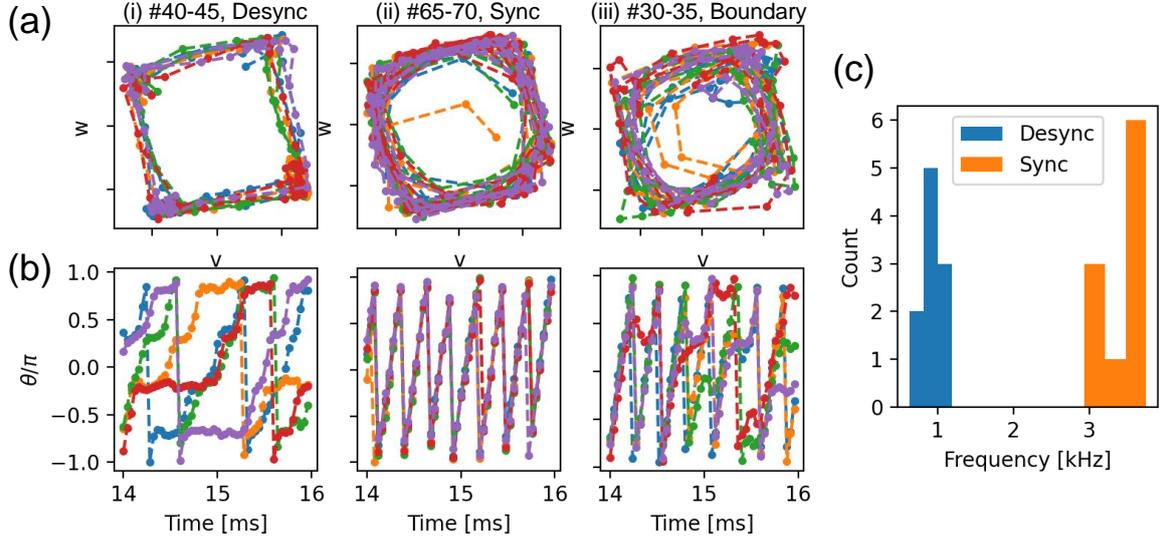

**Figure 3: Class change and mean spiking frequencies** (a) Limit cycle in the v-w plane for sampled neurons in Fig. 2(b) from 12.8 ms to 16.8 ms. (i) neurons #40-45 in desynchronized (Class-I) region, (ii) #65-70 in synchronized (Class-II) region, and (iii) #30-35 at the boundary between regions of two classes. Each color represents a different neuron. (b) Time evolution of rotation phase. (c) Mean spiking frequencies from 14 ms to 16 ms for neurons #40-50 (desynchronized) and #60-70 (synchronized), corresponding to Class-I and Class-II, respectively.

### 3.3 Turbulent chimera state with non-local interaction

Using the same ring structure, we were furthermore able to find a different category of chimera state. Here, we used exponentially decreasing local interactions, as discussed in the Kuramoto model [1]. As shown in Fig. 4a, we set the interaction as $\Gamma(r) = 2^{d_{max}-r-1}$ with $d_{max} = 6$ for $0 < r < d_{max}$ and $\Gamma(r) = 0$ otherwise. The interactions were set to be integer values because of the experimental limitation where $J_{ij}$ has to be a signed 8-bit integer.

Figure 4(b) and (c) show $g_0(t)$ and $h_0$ indices, (d) and (e) show the dynamics of the phases of the neurons $\theta(t)$, and (f) and (g) show the local curvatures $D(t)$ for weak $P/\omega_0 = 0.5$ (left) and strong $P/\omega_0 = 2.6$ (right panels) pump amplitudes. Here, we set $K/\omega_0 = 0.008$ and $0.017$ for weak and strong pump cases, respectively. From Fig. 4(d)-(g), we can see that the regions of desynchronization are not stable for a long time and soon fall back into synchronization. Despite this, these regions tend to reappear out of the synchronized regions. As a result, the local curvature graph shows bubble-like patterns corresponding to neurons falling in and out of synchronization. The category of these chimera states differs from those of the aforementioned ones with constant interactions, where a stationary chimera was observed. As shown in Fig. 4(b) and (c), $g_0(t)$ varies irregularly when $h_0 \sim 0$, indicating that the category is a moving-turbulent chimera. Note that the chimera state with a strong pump (right panels) has a characteristic feature and consists of the synchronized firing region and desynchronized non-firing region. Namely, we again found a spontaneous mode shift and coexistence of different spiking dynamics, firing and non-firing, which is analogous to "chimera death" [30]. This non-firing region corresponds to four stable fixed points resulting from pitchfork bifurcations of $v$- and $w$-DOPOs [42]. As shown in the Supplementary Information, we also found inhomogeneous steady states related to these fixed points [46]. Finally, we should note that antiphase correlation develops in a part of the $D \neq 0$ regions (see the Supplementary Information), suggesting that the observed chimera consists of a mixture of a few types of synchronized and desynchronized regions. A similar characteristic chimera state has been reported in Refs. [24,47] (see the Supplementary Information)..

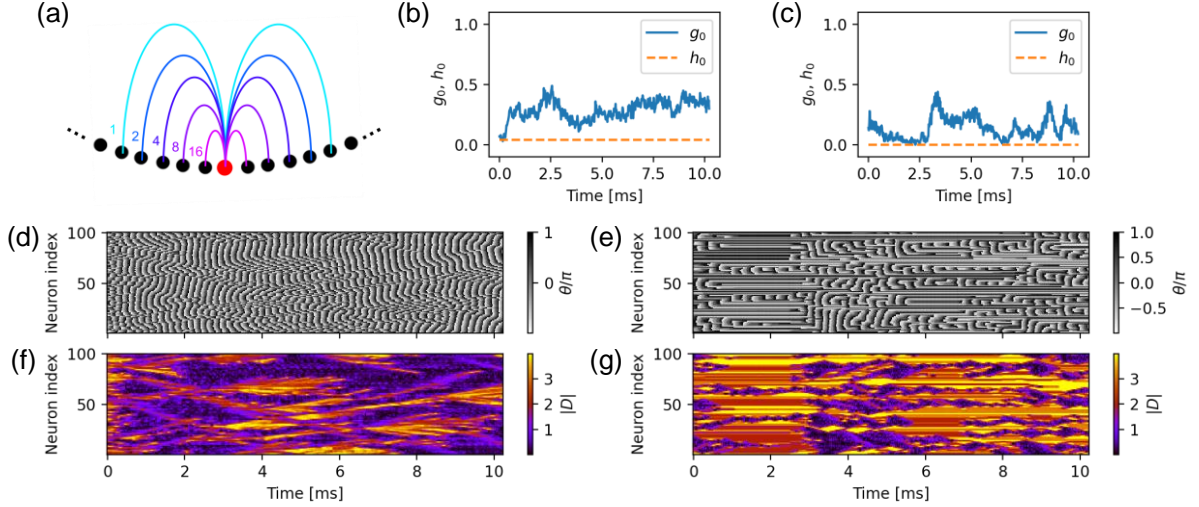

Figure 4: Experimental results for exponential-decay network ($d_{max} = 5$) (a) Network structure of each node. (b) and (c) $g_0(t)$ and $h_0$ indices, (d) and (e) rotation phase of DOPO neurons, (f) and (g) local curvature for weak ($P/\omega_0 = 0.5$) and strong ($P/\omega_0 = 2.6$) pump amplitudes with $K/\omega_0 = 0.008$ and $0.017$, respectively.

## 4. Summary

We demonstrated that a SNN composed of DOPOs can exhibit various chimera states by arranging interactions on a one-dimensional ring lattice. A constant non-local interaction induces a moving-stationary chimera state, which is multi-headed and meandering. An exponentially decaying interaction induces moving-turbulent chimera states, where the desynchronized regions are only stable for short durations and a bubble-like creation-and-annihilation pattern can be observed. The current DOPO neuron has two degrees of freedom, $v$ and $w$, and these variables have inversion symmetry. This means that the fast-slow multiple dynamics in standard SNNs do not exist in the present system, and further that the multiple dynamics are not a prerequisite condition for chimeras. Instead of intrinsic multiple-time-scale dynamics, however, the present system exhibits a spontaneous shift in spiking modes from Class-I to Class-II of Hodgkin's classification. As a result, high-frequency synchronized and low-frequency desynchronized regions appear. This self-organized change in time scale may stabilize the chimera states. Note that the observed chimeras with a high-frequency synchronized region are analogous to neuronal signals observed in Parkinsonian patients with limb tremor [11], while most of the chimeras observed in the previous studies show low-frequency synchronized regions [1,2,14-16, 18,21]. In addition, characteristic meandering movement in the observed moving stationary chimera is induced by this spontaneous change of the spiking modes. We should finally note that the experimental study of chimeras in the SNN is as yet not well established, whereas our system provides an ideal platform for experimental research, since it offers a very high degree of control involving spiking modes, time scales, arbitrary interaction geometry, and so on. The controllability of our system allows us to use the system to perform computations [41] and then to investigate the effects of chimeras on the computation in the future.

## Methods

### Oscillation description of dynamics of neurons

Equations (1) and (2) in the main text can be rewritten by using $v_i + i\, w_i = \sqrt{R_i} \exp i\theta_i$ as follows:

$$\frac{d\theta_i}{dt} = \omega_0 - \frac{R_i}{4}\sin 4\theta_i - \sqrt{\alpha^2 + \beta^2}\sum_j J_{ij} \sin(\theta_i - \theta_j - \phi), \qquad (3)$$

$$\frac{dR_i}{dt} = 2P_i R_i - \frac{R_i^2}{2}(\cos 4\theta_i + 3) + 2\sqrt{\alpha^2 + \beta^2}\, R_i \sum_j \epsilon_{ij} J_{ij} \cos(\theta_i - \theta_j - \phi), \qquad (4)$$

where $\omega_0 = \sqrt{-J_{vw}J_{wv}}$ is the basic spiking frequency of a neuron in the case of zero pump power $P = 0$, $\epsilon_{ij} = R_j/R_i$ and $\phi = \tan^{-1}\frac{\beta}{\alpha}$ with parallel and cross-type coupling strengths $\alpha$ and $\beta$. Equation (3) shows an analogy to the Kuramoto model (and also Stuart Landau model) with exception of the second term on the right-hand side. However, the connection between (3) and (4) induced by the second term causes the characteristic behavior of chimera states discussed in the main text. In the present study, we chose $\phi = \frac{\pi}{2}$ ($\alpha = 1$ and $\beta = 0$), because the Kuramoto model of the ring structure lattice shows chimera states at around $\phi \sim \frac{\pi}{2}$ [1,16].

### Change in spiking frequencies without interneuron couplings

Without any interneuron couplings ($\alpha = 0$ and $\beta = 0$), the change in spiking frequencies is described by the following equation, [41]:

$$\frac{\omega(P)}{\omega_0} = \sqrt{1 - \frac{P^2}{8\omega_0^2}}. \qquad (5)$$

In the limit $P/\omega_0 \to 0_+$, a spiking frequency changes discontinuously from zero to $\omega_0$, which characterizes the Class-II spiking mode. This Class-II mode results from the Andronov-Hopf bifurcation [41]. On the other hand, at $P = \sqrt{8}\omega_0$, the spiking frequency vanishes continuously, which is a characteristic of the Class-I spiking mode. This mode originates from the saddle-node bifurcation [41]. Crossover between these two classes occurs between $P/\omega_0 = 0$ and $\sqrt{8}$.

### Self-organized spiking-mode shifts

As described in the prior paper of Inagaki et al., [41], the order (and hence the synchronization) of the parallel coupled neurons with $\phi = 0$ results in a renormalization of the effective pump power. This can be understood from eq. (3) by considering, for simplicity, a global coupling $J_{ij} = K$, where the term $2R_i K \sum_j \epsilon_{ij} \cos(\theta_i - \theta_j)$ can be rewritten as $2R_i r N K$ by using the order parameter $r = |\sum_j \exp(-i\theta_j)|/N$ with an assumption of $\epsilon_{ij} \sim 1$, and then the term $2P_i R_i$ is renormalized as $2(P_i + rNK)R_i$. Note that synchronized neurons have similar amplitudes, and thus, the assumption $\epsilon_{ij} \sim 1$ is reasonable. This renormalization results in an increase in the effective pump $P_{eff} = P + rNK$ and indirectly decreases the frequency of the oscillators, as described by eq. (5) in the Methods (see also Ref. [41]). In the cross-coupling case, which is used in this experiment, a similar effect occurs. In this case, the effective pump power stays constant, and instead the effective basic frequency $\omega_0$ increases in proportion to the order of the neurons. Namely, when $\phi = \frac{\pi}{2}$, the order parameter can be found in eq. (4), and the renormalization of $\omega_0 + rNK$ can be found from this equation. As a result, we can see that the spiking frequency increases in proportion to the synchronization of the neurons. The renormalization proportional to the order parameter $r$ indicates that this increase/decrease in spiking frequency is spontaneously introduced in a self-organized manner. The Supplementary Information provides experimental evidence of the above discussion.

## Data availability

The data that support the plots within this paper and other findings of this study are available from the corresponding authors upon reasonable request.

## Code availability

The modeling is described in the Methods and Supplementary Information and the code is available from the corresponding authors upon reasonable request.

Networks: Spatial Chaos and Chimera States. *Phys. Rev. Lett.* **106**, 234102 (2011).

## Acknowledgments

This research was partially supported by the Impulsing Paradigm Change through Disruptive Technologies (ImPACT) Program of the Council of Science, Technology and Innovation (Cabinet Office, Government of Japan). T.L. is partially supported by JSPS KAKENHI Grant Number JP22K03545. K.A. is partially supported by the Japan Agency for Medical Research and Development (AMED) under Grant Number JP22dm0307009, the Japan Science and Technology Agency Moonshot R&D Grant Number JPMJMS2021, and JSPS KAKENHI Grant Number JP20H05921. The authors thank Hiroyuki Tamura for his support during this research.

## Author contributions

K. I., T. M., and T. Inagaki proposed the project. T. M. and T. Inagaki performed the experiments. K. I., T. M., and Y. Y. performed the data analysis and numerical simulations. T. L. and K. A. supported the theoretical analysis of the computational neuroscience. T. Inagaki, K. I., T. Ikuta, T. H., and H. T. contributed to build the DOPO network system. K. E., T. U., and R. K. contributed to build the PPLN modules. T. M., K. I., T. Inagaki, Y. Y., T. L., K. A., and H. T. wrote the manuscript with input from all authors.

## Competing financial interests

The authors declare no Competing Non-Financial Interests but the following Competing Financial Interests: K. I., H. T., T. H., T. Inagaki, and T. Ikuta are inventors on patent JP6996457 awarded in December 2021 to NTT that covers a scheme of coupled optical oscillators for SNNs. T. U. and K. E. are inventors on patent JP5856083 awarded in February 2016 to NTT that covers phase-sensitive amplifiers based on periodically poled lithium niobate waveguides. All other authors declare no competing interests

# Supplementary Information

# Experimental observation of chimera states in spiking neural networks based on degenerate optical parametric oscillators

**Authors:** Tumi Makinwa[1,2,†], Kensuke Inaba[1,*,†], Takahiro Inagaki[1,*,†], Yasuhiro Yamada[1], Timothée Leleu[3], Toshimori Honjo[1], Takuya Ikuta[1], Koji Enbutsu[4], Takeshi Umeki[4], Ryoichi Kasahara[4], Kazuyuki Aihara[3] & Hiroki Takesue[1]

[†]These authors contributed equally to this work.

**Affiliations:** [1] NTT Basic Research Laboratories, NTT Corporation, 3-1 Morinosato Wakamiya, Atsugi, Kanagawa, 243-0198, Japan

[2] Department of Applied Sciences, Delft University of technology, Mekelweg 5, 2628 CD Delft, Netherlands

[3] International Research Center for Neurointelligence, The University of Tokyo Institute for Advanced Study, The University of Tokyo, 7-3-1 Hongo, Bunkyo-ku, Tokyo 113-0033, Japan

[4] NTT Device Technology Laboratories, NTT Corporation, 3-1 Morinosato Wakamiya, Atsugi, Kanagawa, 243-0198, Japan

*E-mail to: Kensuke Inaba (kensuke.inaba.yg@hco.ntt.co.jp) and Takahiro Inagaki (takahiro.inagaki.vn@hco.ntt.co.jp)

## S1. Spiking modes and spiking frequencies of a DOPO neuron

The present DOPO neuron shows crossover between a Class-I and Class-II neuron accompanied by a change in spiking frequencies by varying the pump power P [1]. Figure S1 shows typical dynamics of Class-II and Class-I neurons that can be obtained by varying $P/\omega_0$ without any interneuron couplings. We can see that the characteristic of Class-I (II) neurons is a square (circle) trajectory and a stepwise (linear) time dependence of the phase of the neurons $\theta_i = \tan^{-1}\frac{w_i}{v_i}$. The stepwise $\theta_i$ means that the spiking frequencies can continuously decrease to zero by extending the plateau length of the step, which is also a characteristic of Class-I neurons, as mentioned in the Method.

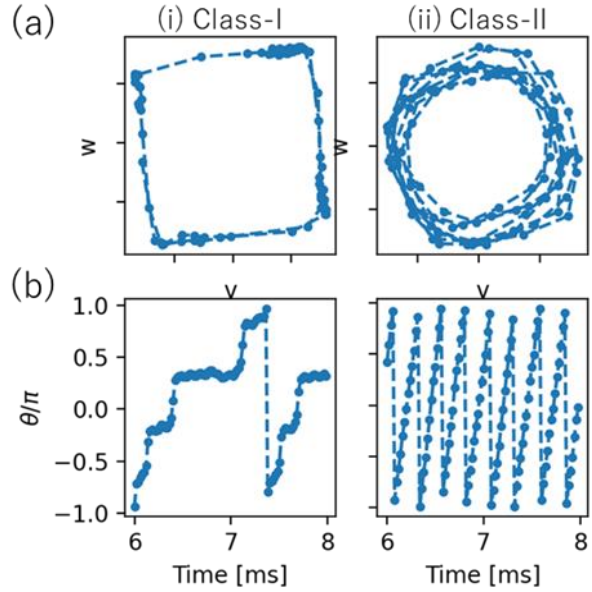

Figure S1: Typical dynamics of Class I (left) and Class-II (right) neurons obtained for large and small pump amplitudes, respectively. Top panels (a) are trajectories in the v-w plane and bottom panels (b) are the time dependence of the neuron phase.

## S2. Adaptive change of spiking frequencies induced within a self-organized manner

As described in the Method, the present system shows a self-organized spiking-mode shift accompanied with a change in spiking frequency. Figure S2 shows experimental evidence of this change in spiking frequency. Here, we set all-to-all connections in each group of ten neurons, and varied the coupling strength and coupling type (parallel or cross). The cross-type coupling shows an increase in the spiking frequencies as the coupling strength increases, suggesting a crossover from Class-I to Class-II, while the parallel type instead shows a decrease in the spiking frequencies, suggesting a crossover from Class-II to Class-I.

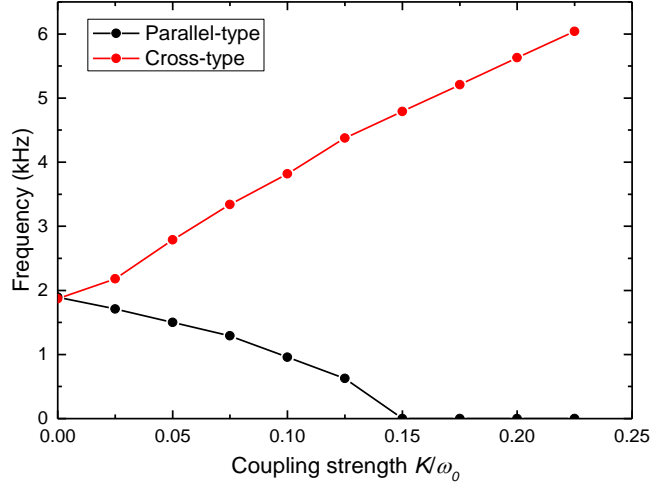

Fig. S2: Change in the spiking frequency caused by synchronization. Each neuron in a group of ten neurons interacts with each other, and the coupling strengths are set to different values.

## S3. Alternating feature of the moving stationary chimera

The moving stationary chimera observed in the constant interactions shows characteristic meandering movement as discussed in the main text. This meandering movement shows an alternating chimera-like feature [2,3], which has been discussed in relation to unisemispheric sleep [4]. Figure S3 shows the order parameter for neurons #55-60 (top) and #30-35 (bottom). In these regions, the order parameters alternately increase and decrease. Or, in the other words, the synchronized and desynchronized regions alternate.

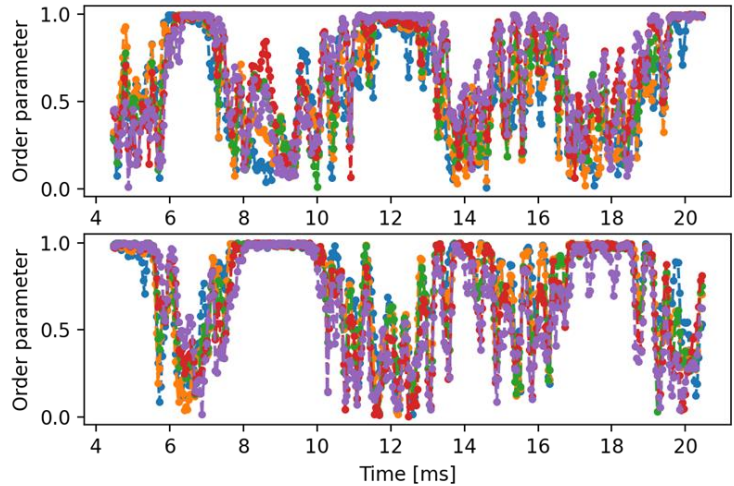

Figure S3: Time dependence of the order parameters of neuron indices #55-60 (top) and #30-35 (bottom). The experimental parameters are the same as in Fig.2 and 3 in the main text.

## S4. Numerical simulation of meandering chimera

To investigate the effects of experimental noise on the meandering movement of the chimera state, we performed a numerical simulation based on eqs. (1) and (2) in the main text. Figure S4 shows the dynamics of the phases and the local curvatures of the neurons. Without noise, the numerical simulation still reproduces the meandering movement, suggesting that such a movement does not come from experimental noise. Note that the initial states in numerical simulation are randomly given.

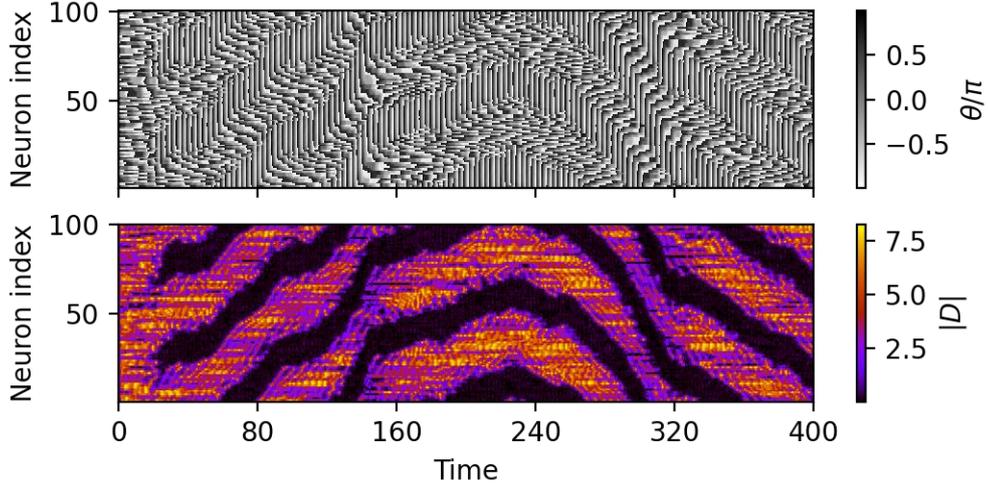

Figure S4: Numerical simulation results for $P/\omega_0 = 2.4$ and $K/\omega_0 = 0.067$. (a) Time dependence of $\theta_i$ and (b) local curvature of each neuron.

## S5. Coherent and incoherent examples

We also observed almost fully synchronized coherent and desynchronized incoherent states. Figure S5(a) and (b) show the dynamics of the phases of the neurons $\theta(t)$ and the local curvatures $D(t)$, and (c) show the density of the synchronized component $g_0(t)$ and that of the temporary stable component $h_0$ [5], where we have set a small pump amplitude $P/\omega_0 = 0.5$ and interaction strength $K/\omega_0 = 0.033$. Figure S5 (a) and (b) show a coherent state that appears after a transient time of about 0.5 ms. Here $g_0(t)$ and $h_0$ stay near 1, suggesting a synchronized state, too. Note that experimental fluctuations of the optical amplitudes cause the noisy $g_0(t)$.

Figure S6 shows the same quantities as shown above in Fig. S5 for a large pump amplitude $P/\omega_0 = 2.6$. Here, we found that $g_0(t)$ goes to zero, suggesting that an asynchronous incoherent state appears. We found that a transient chimera state remains until around 5 ms. After that, all of the neurons fall into stable fixed points related to pitchfork bifurcations, where four fixed points are stable due to the symmetric $v$ and $w$ components [1]. Note that the obtained steady state is inhomogeneous [6,7]. Antiphase correlations can be found in some parts of this desynchronized region [8]. These results strongly suggest the relevance for the chimera-death-like state [9], as discussed in the main text.

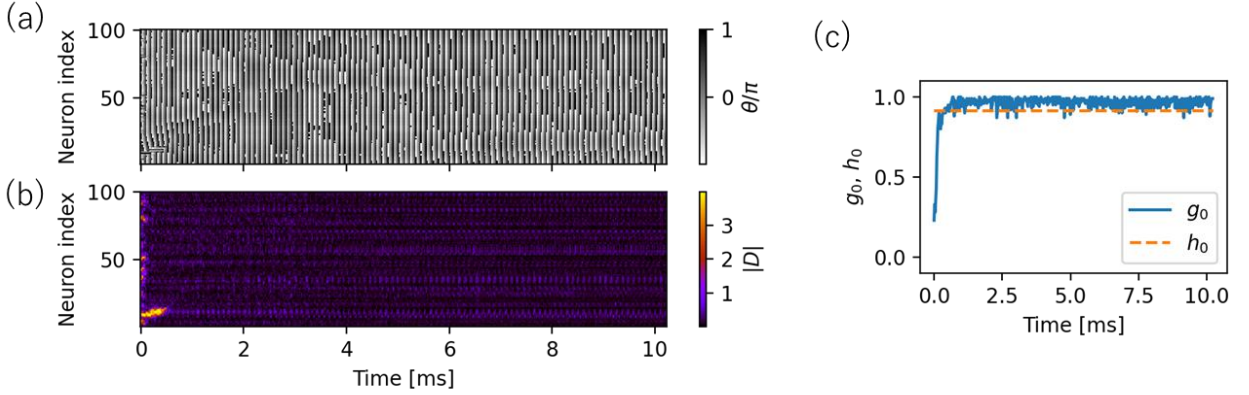

Figure S5: (a) Dynamics of the neuron phase $\theta_i$ (b) local curvature $|D|$, and (c) density of coherent component $g_0(t)$ and temporary stable component $h_0$ for a small pump amplitude $P/\omega_0 = 0.5$ and interaction strength $K/\omega_0 = 0.033$.

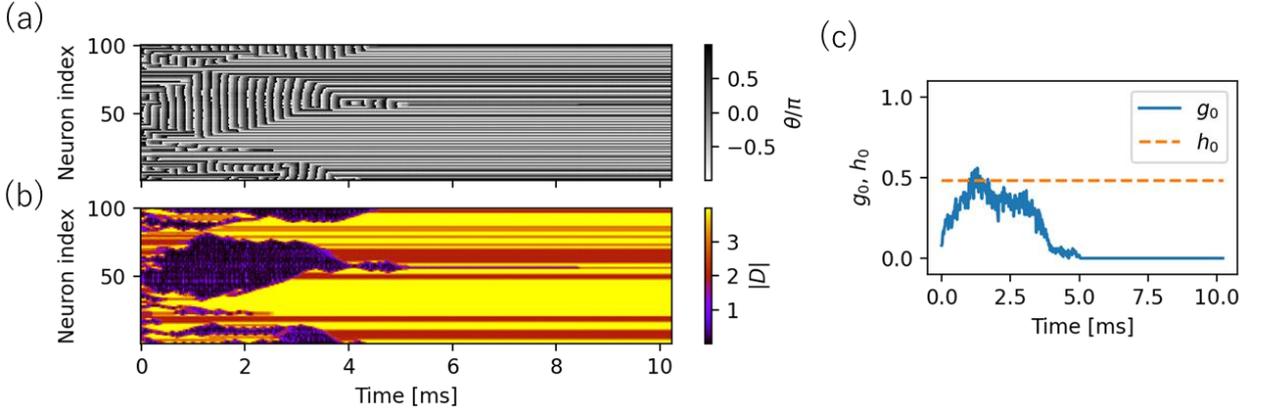

Figure S6: (a) Dynamics of the neuron phase $\theta_i$, (b) local curvature $|D|$, and (c) density of coherent component $g_0(t)$ and temporary stable component $h_0$ for a large pump amplitude $P/\omega_0 = 2.6$ and interaction strength $K/\omega_0 = 0.033$.

## S6. Antiphase coherence

Figure S7 (a) and (b) show the cluster order parameters $r_k = |\sum_{j \in C_k} \exp(-i\theta_j)|/N_C$ and antiphase ($\pi$-phase) cluster order parameters $r_{\pi,k} = |\sum_{j \in C_k} \exp(-i\theta_j - i\pi j)|/N_C$ for the same experimental parameters as Fig. 4 (b), (d), and (f) in the main text, where $C_k = \{k - \frac{d_{\max}}{2}, k -$

$\frac{d_{max}}{2} + 1, \dots, k + \frac{d_{max}}{2}\}$ and $N_C = d_{max} + 1$. We can extend these quantities to $\pm \pi/2$-phase correlations as $r_{\pm\pi/2,k} = \left|\sum_{j \in C_k} \exp\left(-i\theta_j \mp i\frac{\pi}{2}j\right)\right|/N_C$, as shown in Fig. S7(c) and (d). Figure S7(e)-(h) show these order parameters for the experimental parameters in Figs. 4 (c), (e), and (g) of the main text. As discussed in, e.g., Refs. [10,11], the cluster order parameter $r_k$ characterizes the phase coherence of each neuron. The $\pi$-phase and $\pm\pi/2$-phase order parameter characterizes an antiphase (two-site-period) and four-site-period-phase coherence in the same manner. From Figs. S7 (b)-(d) and (f)-(g), we found that parts of the region with a finite local curvature $D$ have large magnitudes, $r_\pi \sim 1$ or $r_{\pm\pi/2} \sim 1$, suggesting that there is an antiphase ($\pi$-phase) or $\pm\pi/2$-phase coherence. Enhancements of these $\pi$- and $\pm\pi/2$-phase coherences are related to the four stable fixed points, as mentioned in Sec. S5. We concluded that the chimera states obtained here consist of a mixture of different regions, including a coherent synchronized region ($D \sim 0$ and $r \sim 1$), an antiphase coherent synchronized region ($D \sim D_{max}$ and $r_\pi \sim 1$), an $\pm\pi/2$-phase coherent synchronized region n ($0 < D < D_{max}$ and $r_{\pm\pi/2} \sim 1$), and an incoherent desynchronized region ($0 < D < D_{max}$ and $r, r_\pi, r_{\pm\pi/2} < 1$). A similar chimera state consisting of two different coherent regions, i.e., two (normally) synchronized regions with largely different amplitudes, and an incoherent region has been recently reported [10]. The period doubling transition in a chaotic system also induces similar chimera states, as discussed in Ref. [8].

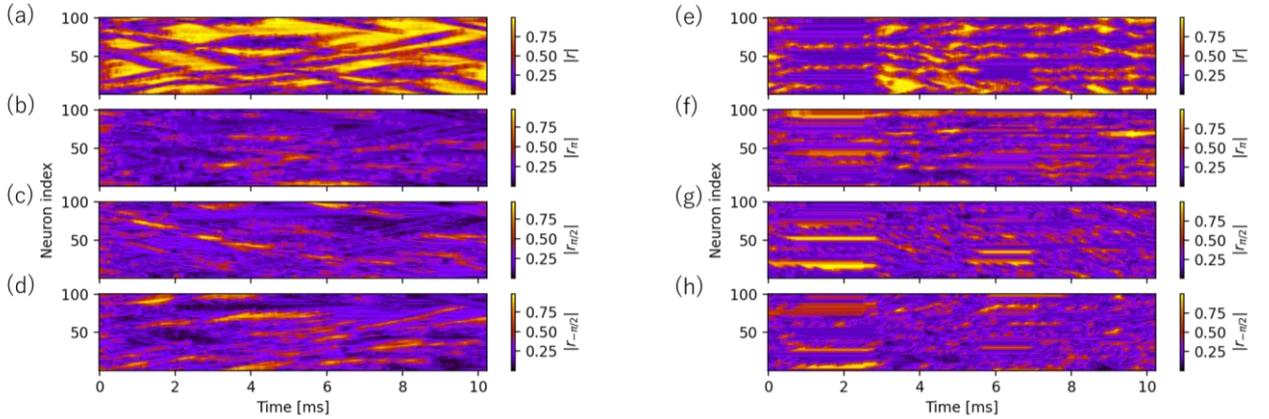

Figure S7: Order parameter (a), antiphase ($\pi$-phase) (b), $\pi/2$-phase and $-\pi/2$-phase order parameters (c) and (d) for the same experimental parameters as in Fig. 4(b), (d), and (f) of the main text. (e)-(h) The same order parameters for the experimental parameters in Fig. 4(c), (e), and (g) of the main text.

## S7. Time evolutions of optical amplitudes

Here, we show the amplitudes $\sqrt{v_i^2 + w_i^2}$ of the data presented in the main text and in this note. We can see that the synchronized and desynchronized phases have different amplitudes from each other, suggesting a feature like found in the amplitude chimera states [7].

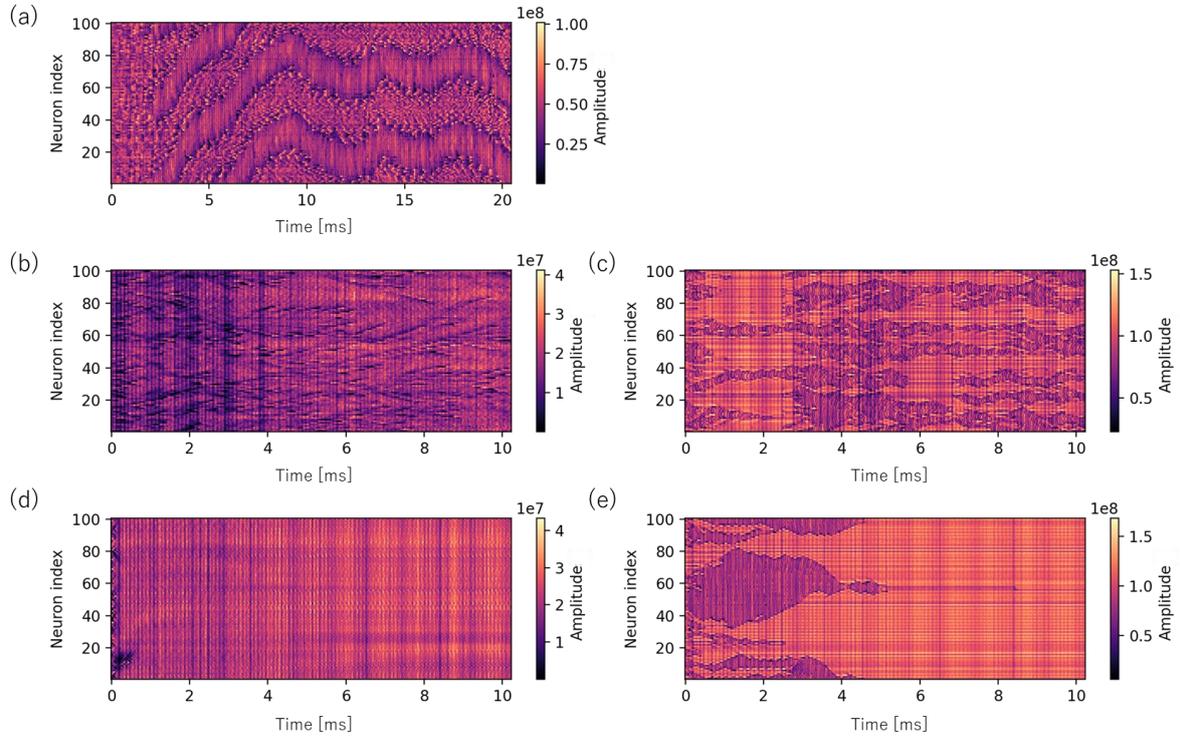

Fig. S8: Amplitudes of DOPOs, $\sqrt{v_i^2 + w_i^2}$ for stationary double headed chimera (a), turbulent chimeras (b) and (c) in the main text, and coherent and inhomogeneous steady states in this note (d) and (e).